\documentclass{sn-jnl} 
\usepackage{graphicx}%
\usepackage{multirow}%
\usepackage{amsmath,amssymb,amsfonts}%
\usepackage{amsthm}%
\usepackage{mathrsfs}%
\usepackage[title]{appendix}%
\usepackage{xcolor}%
\usepackage{textcomp}%
\usepackage{manyfoot}%
\usepackage{caption}%
\usepackage{booktabs}%
\usepackage{algorithm}%
\usepackage{algorithmicx}%
\usepackage{hyperref}%
\usepackage{algpseudocode}%
\usepackage{listings}%
\usepackage[12pt]{extsizes}
\usepackage{adjustbox}
\usepackage{geometry}
\theoremstyle{thmstyleone}%

\theoremstyle{thmstyletwo}%

\theoremstyle{thmstylethree}%

\begin{document}

\title[Article Title]{Using Internal Bar Strength as a Key Indicator for Trading Country ETFs}

\author[1]{\fnm{Aditya} \sur{Pandey}}\email{adityapandey@nyu.edu}
\equalcont{These authors contributed equally to this work.}

\author[1]{\fnm{Kunal} \sur{Joshi}}\email{kjoshi@nyu.edu}
\equalcont{These authors contributed equally to this work.}

\affil[1]{\orgdiv{Stern School of Business}, \orgname{New York University}, \orgaddress{\city{New York}, \postcode{10003}, \state{NY}}}

\abstract{In this report, we aim to investigate the effectiveness of using internal bar strength (IBS) as a key indicator for trading country exchange-traded funds (ETFs). The study uses a quantitative approach to analyze historical price data for a bucket of country ETFs over a period of 10 years and use the idea of Mean Reversion to create a profitable trading strategy. 
Our findings suggest that IBS can be a useful technical indicator for  predicting short-term price movements in this basket of ETFs.}


\maketitle

\section{Introduction}

The Internal Bar Strength Indicator measures the position current trading session's close relative to the session's high-low range. It is generally used as an indicator of mean-reversion.

Pagonidis \cite{effect} proposed that stocks that close with an IBS below 0.2, will rise in price the next day, while stocks that close with an IBS above 0.8, will often decline in value in the following session.

These findings were on daily data of various ETFs covering the period from the inception for each ETF to 5/12/2013. We tested this on more recent data to find similar results.
Our approach is similar to the above mentioned one in which if the IBS of a country ETF is low, it suggests that the price has closed near the low of its daily range, which may indicate that the market is oversold and due for a bounce back up. This could be an opportunity to enter a long position with the expectation that the price will rise. 

Conversely, if the IBS of a country ETF is high, it suggests that the price has closed near the high of its daily range, which may indicate that the market is overbought and due for a correction or pullback. This could be an opportunity to enter a short position with the expectation that the price will fall.

We present a strategy based on this that aims to be always in and hedge across multiple ETFs. We also present modifications to this strategy that show how changes in holding periods and number of ETFs invested in affects the performance of the strategy.

\section{Data and Methodology}

\subsection{Data}

Our analysis uses ETF data from 2009/1/1 to 2019/12/31. ETF data is sourced from Yahoo Finance \cite{yahoo}. The descriptions are in \textbf{Table \ref{ETFs}.}
\begin{table}[htpb]
\caption{ETFs}
\label{ETFs}
\large
\begin{tabular}{|c|c|c|}
\hline
\textbf{Country} & \textbf{Ticker} & \textbf{ETF}                     \\ \hline \hline
India            & PIN             & Invesco India ETF                \\ \hline
China            & FXI             & iShares MSCI China Large-Cap ETF \\ \hline
South Korea      & EWI             & iShares MSCI Italy ETF           \\ \hline
Mexico           & EWW             & iShares MSCI Mexico ETF          \\ \hline
South Africa     & EZA             & iShares MSCI South Africa ETF    \\ \hline
Taiwan           & EWT             & iShares MSCI Taiwan ETF          \\ \hline
Japan            & EWJ             & iShares MSCI Japan ETF           \\ \hline
USA              & IVV             & iShares Core S\&P 500 ETF        \\ \hline
UK               & EWU             & iShares MSCI United Kingdom ETF  \\ \hline
EU               & EZU             & iShares MSCI Eurozone ETF        \\ \hline
Australia        & EWA             & iShares MSCI Australia ETF       \\ \hline
Singapore        & EWS             & iShares MSCI Singapore ETF       \\ \hline
Canada           & EWC             & iShares MSCI Canada ETF          \\ \hline
Israel           & EIS             & iShares MSCI Israel ETF          \\ \hline
Brazil           & EWZ             & iShares MSCI Brazil ETF          \\ \hline
\end{tabular}
\end{table}

\subsection{Methodology}

IBS over $n$ days is calculated as:
\begin{equation}
    IBS_n = \frac{Close - Low_n}{High_n - Low_n}
\end{equation}
where the $Low_n$ and $High_n$ could be calculated over $n$-days ($n=1$ being the default)
\\
\\
$IBS$ takes values from 0 (close is lowest) to 1 (close is highest).
We analysed our data in Python using Numpy and Pandas. Charts were created using Seaborn.

\pagebreak

\section{The Strategy}

The threshold strategy is used on individual ETFs and has varying performance. We show this performance in \textbf{Table \ref{single_vs_minmax}}. 

\begin{table}[htpb]
\caption{Sharpe Ratios of Min-Max IBS Strategy and Single ETF Threshold Strategy} 
\label{single_vs_minmax}
\begin{tabular}{|c|c|c|}
\hline
\textbf{Strategy/ETF} & \textbf{Sharpe Ratio} & \textbf{Time In} \\ \hline
IBS Min-Max            & \textbf{2.907858}         & \textbf{100\%}    \\ \hline
PIN                    & 2.166918         & 54.463\% \\ \hline
EWJ                    & 1.585597         & 41.489\% \\ \hline
EWI                    & 1.475513         & 46.91\%  \\ \hline
EIS                    & 1.259877         & 47.488\% \\ \hline
EZA                    & 1.080578         & 45.573\% \\ \hline
FXI                    & 1.051496         & 39.357\% \\ \hline
EWA                    & 0.974023         & 45.356\% \\ \hline
EWT                    & 0.956515         & 42.031\% \\ \hline
EZU                    & 0.708747         & 48.103\% \\ \hline
EWS                    & 0.665158         & 43.188\% \\ \hline
EWZ                    & 0.482336         & 44.958\% \\ \hline
EWU                    & 0.288471         & 43.802\% \\ \hline
IVV                    & 0.200902         & 48.066\% \\ \hline
EWW                    & -0.192655        & 46.765\% \\ \hline
EWC                    & -0.451279        & 44.814\% \\ \hline
\end{tabular}
\end{table}

However, the time in for these strategies is low which hurts the Sharpe Ratio. Our strategy takes a basket of multiple ETFs and considers daily IBS values for all.

We also include a the probabilities of getting a positive return on following a threshold bases strategy in \textbf{Table \ref{positive_prob}}. These probabilities are similar to the ones seen in \cite{effect} for the probability of an Up day split by IBS quintiles.

\begin{table}[htpb]
\caption{Probabilities of Positive Returns} 
\label{positive_prob}
\begin{tabular}{|c|c|c|}
\hline
\textbf{Ticker} & \textbf{Long on IBS \textless 0.2} & \textbf{Short on IBS \textgreater 0.8 }\\ \hline
EWJ    & 0.606061                  & 0.514364                      \\ \hline
EIS    & 0.587719                  & 0.513981                      \\ \hline
PIN    & 0.580692                  & 0.578240                      \\ \hline
EWT    & 0.572383                  & 0.502770                      \\ \hline
FXI    & 0.567686                  & 0.523052                      \\ \hline
IVV    & 0.565502                  & 0.460481                      \\ \hline
EZU    & 0.552209                  & 0.476647                      \\ \hline
EWS    & 0.550308                  & 0.514563                      \\ \hline
EWI    & 0.548000                  & 0.511222                      \\ \hline
EWZ    & 0.543672                  & 0.508006                      \\ \hline
EZA    & 0.540835                  & 0.509859                      \\ \hline
\end{tabular}
\end{table}

We see that the distribution of minimum and maximum IBS values is heavy-tailed. The minimum values skew towards 0 and the maximum values skew towards 1.

\begin{figure}[H]
\caption{Distribution of Min and Max IBS values}
    \centering
    \includegraphics[scale=0.7]{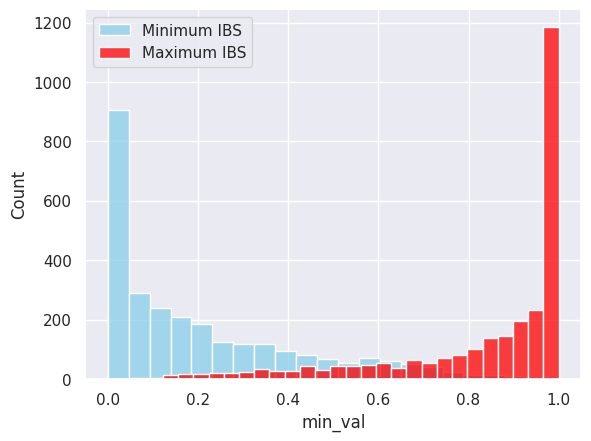}
\end{figure}

We then implement a \textbf{min-max} strategy which goes \textbf{long on the ETF which has the minimum IBS} of that day and goes \textbf{short on the ETF which has the maximum IBS} of that day.

We enter at close of the current day and exit at the close of the next day. This gives us much better results overall. We have included a snippet of our strategy performance when we randomize the basket size and choice of ETFs in the basket.

\begin{table}[htpb]
\caption{Best Performing Buckets} 
\label{best_buckets}
\begin{tabular}{|c|c|c|c|c|}
\hline
\textbf{ETF Basket} & \textbf{SR} & \textbf{Long Only SR} & \textbf{Short Only SR} & \textbf{N (No. of ETFs)} \\ \hline
india, tw, can, israel, uk, spore, aus & 3.904 & 1.705 & 0.839 & 2 \\ \hline
india, tw, can, israel, uk, spore, aus & 3.733 & 1.997 & 1.246 & 1 \\ \hline
israel, tw, japan, spore, sk, india      & 3.424 & 1.541 & 0.968 & 2 \\ \hline
spore, india, usa, aus, israel, uk       & 3.354 & 1.562 & 0.621 & 2 \\ \hline
uk, india, sk, spore, china, eu          & 3.311 & 1.475 & 1.200 & 1 \\ \hline
china, uk, india, sk, spore, can         & 3.279 & 1.472 & 1.234 & 1 \\ \hline
china, uk, india, sk, spore, can         & 3.201 & 1.310 & 0.772 & 2 \\ \hline
uk, india, sk, spore, china, eu          & 3.184 & 1.257 & 0.791 & 2 \\ \hline
spore, india, usa, aus, israel, uk       & 3.153 & 1.742 & 1.015 & 1 \\ \hline
brazil, aus, tw, china, uk, sa, india  & 3.089 & 1.338 & 0.724 & 2 \\ \hline
\end{tabular}
\end{table}

In our opinion, choosing from a basket of country ETFs based on their IBS offers advantages compared to just using a single pair of ETFs.

Using a basket of ETFs can provide diversification benefits by spreading the risk across multiple countries, rather than constantly re-using the same pair. This helps reduce the overall risk. 

Using a basket of ETFs also allows for more opportunities for trading. With multiple countries in the basket, there are more frequent opportunities for trades based on IBS signals as there is a much higher probability of having the $min(IBS)$ and $max(IBS)$ of the ETFs in the desired thresholds or having these values sufficiently low/high.

\pagebreak

\section{Results}

\subsection{Comparing the performance to buy and hold strategy}

The strategy significantly outperforms a buy and hold strategy for all ETFs that we considered.

\begin{figure}[H]
    \centering
    \caption{Equity graphs comparing MinMax Strategy to Buy-N-Hold Strategies}
    \includegraphics[scale=0.7]{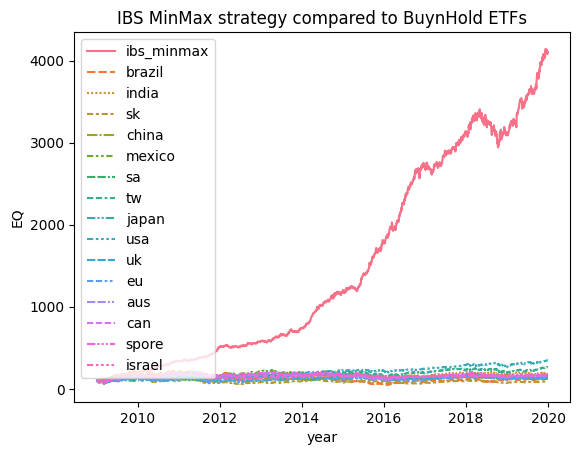}
\end{figure}

\subsection{Comparing the performance across basket sizes}

We ran our strategy on random combinations ranging from size 2 to size 14 of the ETFs in \textbf{Table \ref{ETFs}}. We see a trend of increasing Sharpe ratios with an increase in basket size.

\begin{figure}[H]
    \centering
    \caption{MinMax Strategy performance across basket sizes}
    \includegraphics[scale=0.7]{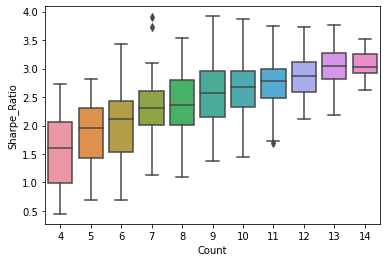}
\end{figure}

This can be alluded to a higher likelihood of the minimum and maximum IBS values being closer to 0 and 1 respectively (i.e. if .


\subsection{Comparing the performance between long only and short only strategies}

We see that long only strategies have a relatively higher Sharpe Ratio than short only strategies (see \textbf{Table \ref{best_buckets}}) and generally tend to perform better equity graphs.

\begin{figure}[H]
    \centering
    \caption{Equity Graphs for Long Only Strategy across top-10 buckets}
\includegraphics[scale=0.4]{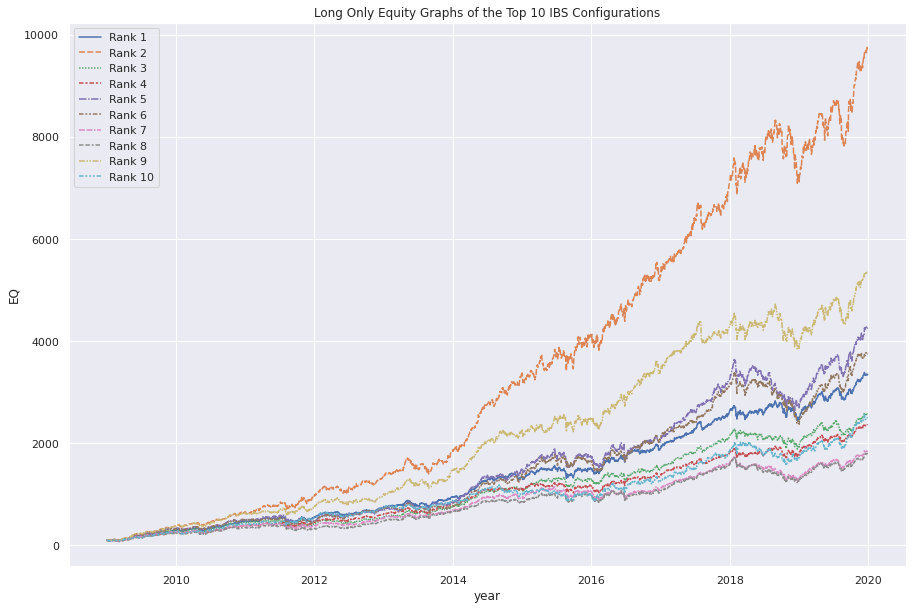} \\
\end{figure}

\begin{figure}[H]
    \centering
    \caption{Equity Graphs for Short Only Strategy across top-10 buckets}
\includegraphics[scale=0.4]{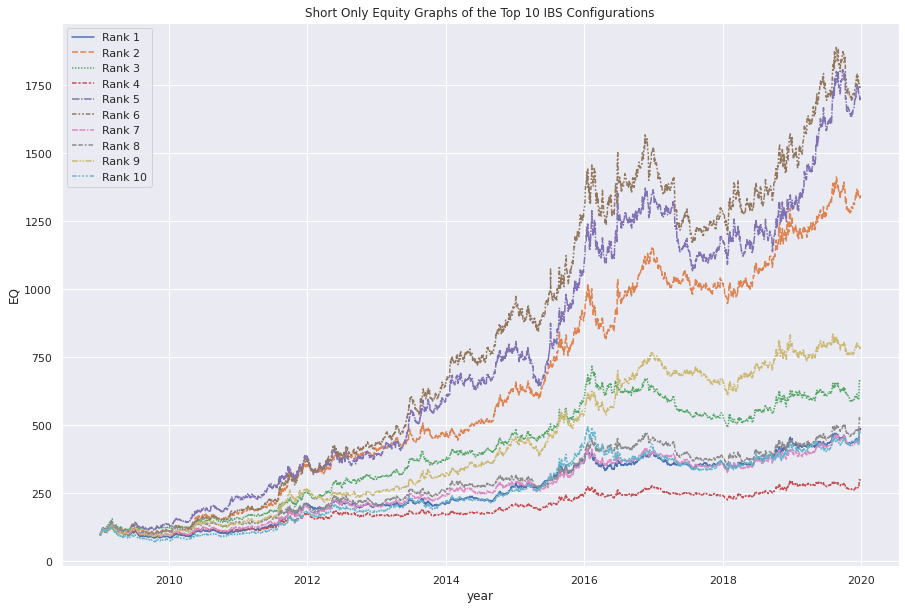}
\end{figure}

\subsection{Comparing the performance across multi-ETF trades}

We modify our strategy slightly by sorting the IBS values and pick the \textbf{top-N} and \textbf{bottom-N} ETFs to go short and long on respectively.
\begin{figure}[H]
    \centering
    \caption{Strategy Performance across number of ETFs held}
    \includegraphics[scale=0.5]{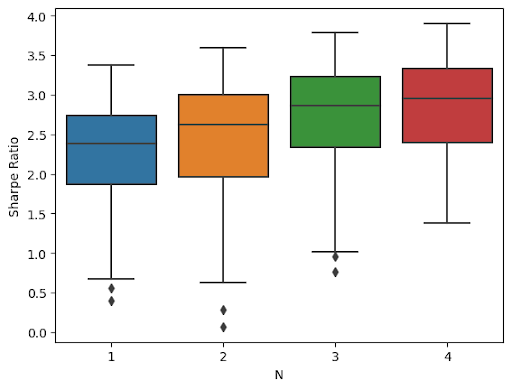}
\end{figure}

An increase in \textbf{N}, increases the Sharpe ratio. As we go long (or short) on multiple ETFs we hedge against an ETF that performs against expectations. 

\subsection{Comparing the performance across multi-day holding periods}

All the previous strategies were over a 1-day holding period. The strategy below varies the holding period between 1 and 4 days.

\begin{figure}[H]
    \centering
    \caption{Strategy Performance across Holding Periods and ETFs held}
    \includegraphics[scale=0.5]{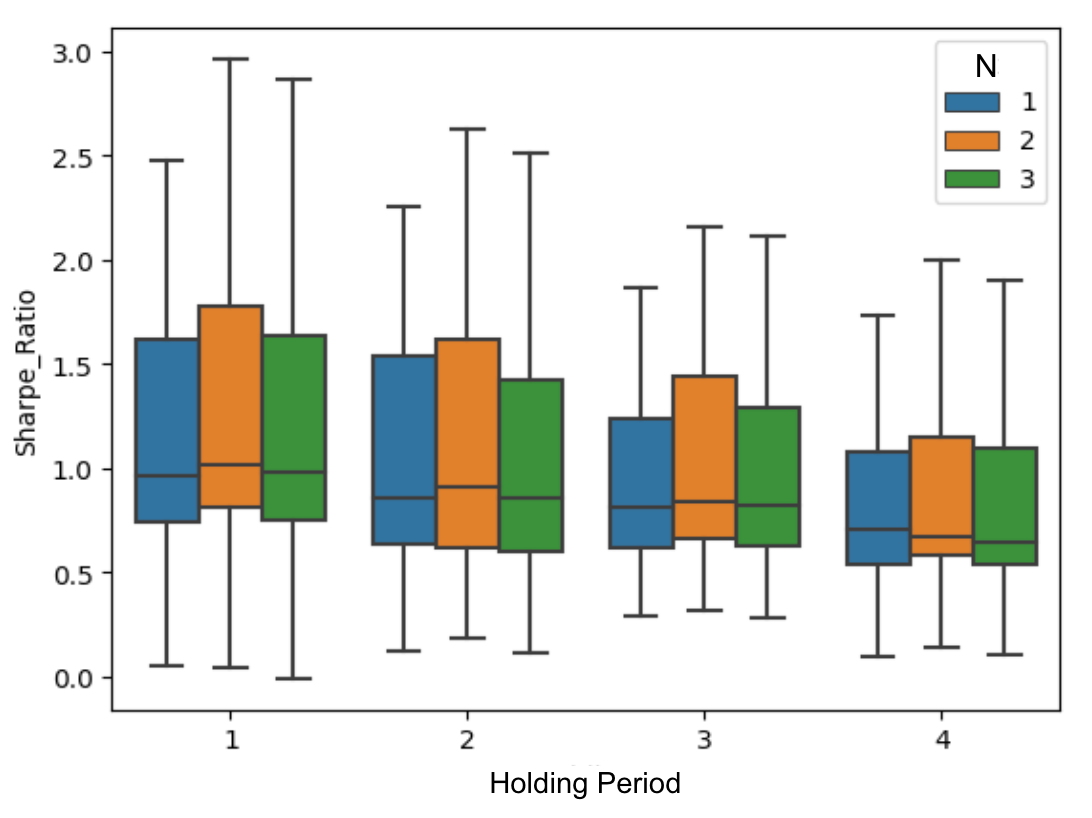}
\end{figure}

We see a general drop in performance as holding period increases. The 1 day IBS indicator does not seem to have an effect beyond horizons of one day.


\subsection{Comparing the performance across multi-day IBS calculations}

We tried to calculated IBS over a 2-day period. This strategy also sees a drop in performance compared to a 1-day IBS metric.
\begin{figure}[H]
    \centering
    \caption{IBS Calculation Period vs. Sharpe Ratio}
    \includegraphics[scale=0.8]{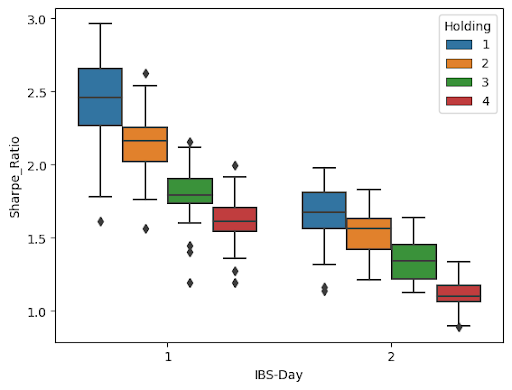}
\end{figure}


\subsection{Comparing Close-to-Close MinMax with Open-to-Open MinMax}

All the strategies so far have been Close-to-Close trades. We compare this with an Open-to-Open strategy to see the importance of getting in at Close.

\begin{table}[htpb]
\caption{Open-Open vs Close-Close Performance of our Strategy} 
\label{openvsclose}
\begin{tabular}{|c|c|c|}
\hline
\textbf{ETF\_Basket} & \textbf{Close to Close} & \textbf{Open To Open} \\ \hline
\textbf{india, tw, can, israel, uk, spore, aus}       & 3.733541                & -0.250662             \\ \hline
\textbf{uk, india, sk, spore, china, eu}              & 3.311943                & -0.011744             \\ \hline
\textbf{china, uk, india, sk, spore, can}             & 3.279045                & 0.286050              \\ \hline
\textbf{spore, india, usa, aus, israel, uk}           & 3.153894                & -0.269298             \\ \hline
\textbf{brazil, india, sa, japan, tw, spore}          & 2.873541                & 0.262017              \\ \hline
\textbf{aus, brazil, sa, india, eu, tw, spore}        & 2.869931                & 0.181264              \\ \hline
\textbf{sa, can, china, uk, india}                    & 2.822592                & 0.251654              \\ \hline
\textbf{brazil, aus, tw, china, uk, sa, india}        & 2.789485                & 0.142810              \\ \hline
\textbf{israel, tw, japan, spore, sk, india}          & 2.735891                & -0.124966             \\ \hline
\textbf{brazil, spore, can, sa, china, india, mexico} & 2.733041                & 0.362022              \\ \hline
\end{tabular}
\end{table}

As seen from \textbf{Table \ref{openvsclose}} the Sharpe Ratios for a Close-to-Close strategy are not good underlying the \textbf{importance of getting in at close}.

\pagebreak

\subsection{Comparing MinMax strategy to a Threshold based strategy over ETF baskets}

We go back to the threshold strategy. Instead of implementing it on a single ETF, we implement it over our ETF baskets. It goes long and short on ETFs if their respective IBS values exceed predefined thresholds. \\
\textbf{Note}: The strategy does not enter a trade unless both long and short thresholds are crossed.
\\
\begin{figure}[H]
    \centering
    \caption{Threshold and MinMax strategies over number of ETFs held (1-Day Holding)}
    \includegraphics[scale=0.6]{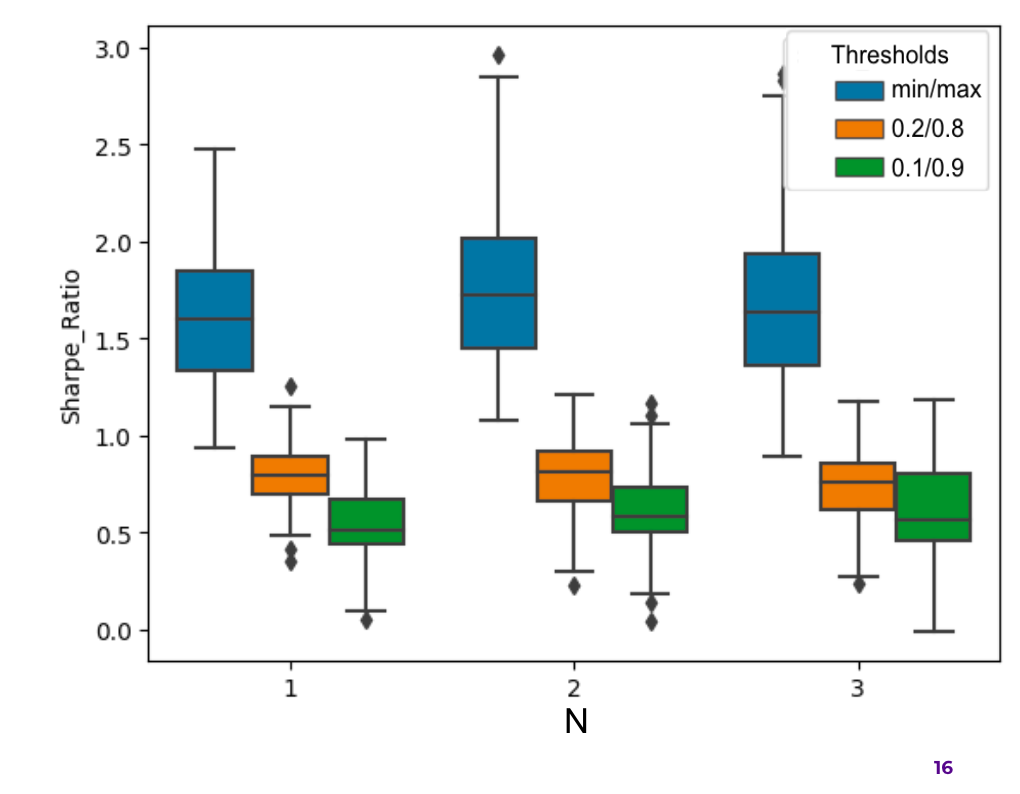}
\end{figure}


\begin{figure}[H]
    \centering
    \caption{Threshold and MinMax strategies over holding periods (Pair-ETFs)}
    \includegraphics[scale=0.6]{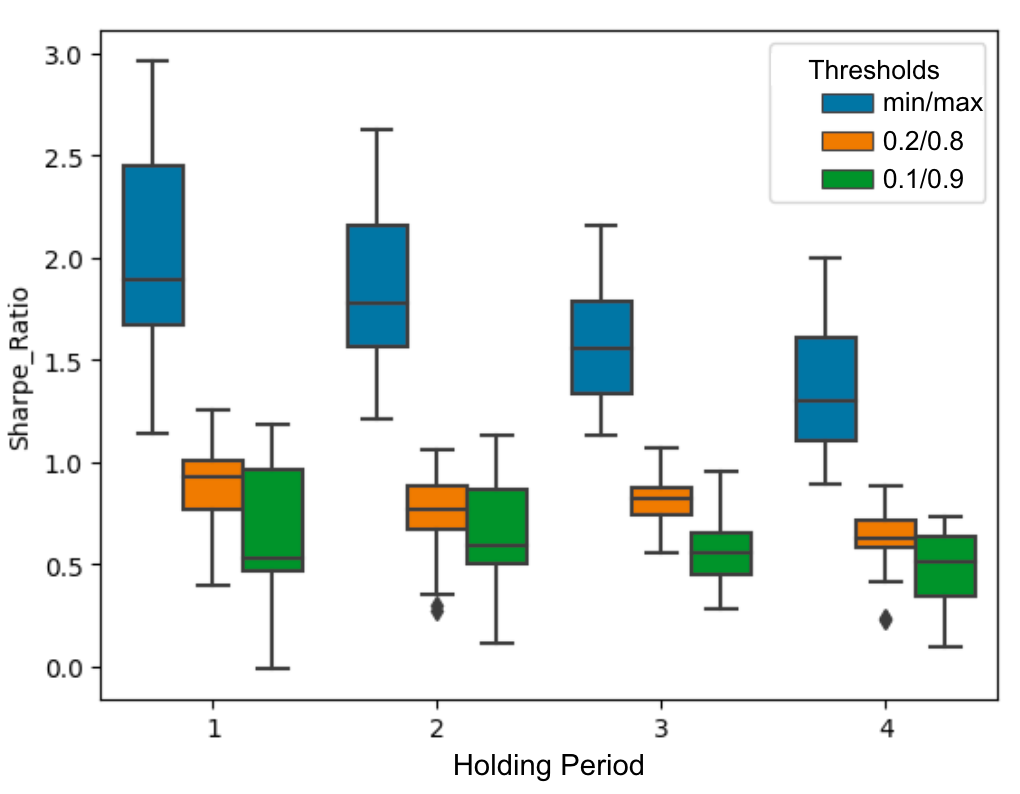}  
\end{figure}

We compare the above strategy with our MinMax strategy across holding periods as well as multi-ETF trades. The threshold strategies universally have a worse performance than the MinMax strategies.

\section{Effect of Trading Costs and Slippage}

IBS is calculated at the Close. Our strategy also enters at the Close price. In practice, this is difficult to setup. Realistically, we can calculate the IBS taking the price just before close and then entering (or not) right after. This will result in slippage and may affect performance

Shorting ETFs also has an additional borrow cost. We have assumed a borrow rate of 0.01\% daily. However, this rate is not constant and varies across ETFs. We performed an analysis which compares strategy performs across a range of interest rates which can be seen in \textbf{Table \ref{borrowing_costs}}. The strategy performs reasonably well till a borrow rate of around 0.15\% daily (approx. 55.75\% annual) after which borrow costs negate performance.

In situations where borrow rates are high, a long-only strategy would still perform well (See \textbf{Table \ref{best_buckets}}).
\\
\begin{figure}[H]
    \centering
    \caption{Borrow Rate (in \%) vs. Sharpe Ratio}
    \includegraphics[scale=0.45]{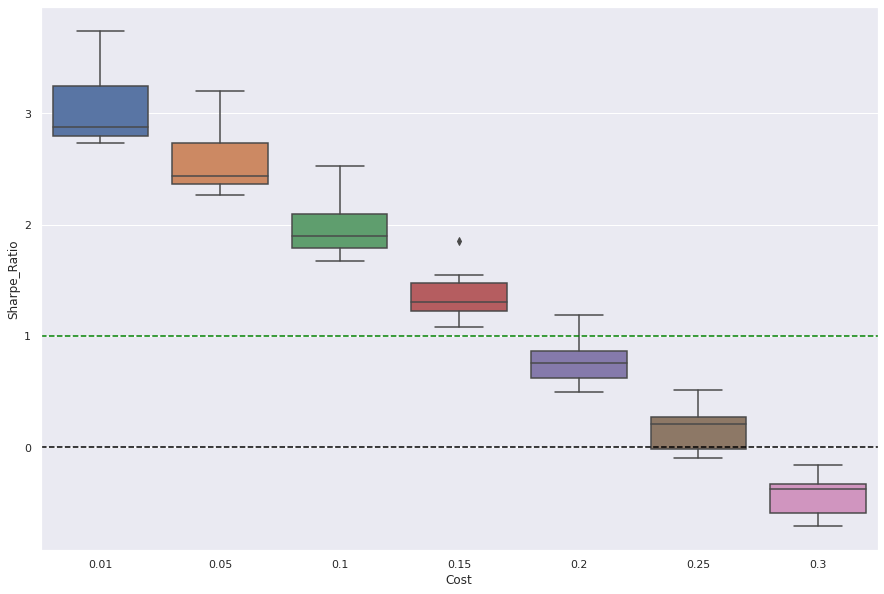}    
\end{figure}

\begin{table}[htpb]
\caption{How Borrowing Costs affect Sharpe Ratio} 
\label{borrowing_costs}
\begin{tabular}{|c|c|c|c|c|c|c|c|}
\hline
\textbf{ETF Basket vs Daily Borrowing Costs} &
  \textbf{0.01\%} &
  \textbf{0.05\%} &
  \textbf{0.1\%} &
  \textbf{0.15\%} &
  \textbf{0.2\%} &
  \textbf{0.25\%} &
  \textbf{0.3\%} \\ \hline
\textbf{india, tw, can, israel, uk, spore, aus} & 3.733 & 3.197 & 2.526 & 1.856 & 1.185 & 0.515  & -0.155 \\ \hline
\textbf{uk, india, sk, spore, china, eu}          & 3.311 & 2.808 & 2.178 & 1.548 & 0.919 & 0.289  & -0.339 \\ \hline
\textbf{china, uk, india, sk, spore, can}         & 3.279 & 2.770 & 2.134 & 1.499 & 0.863 & 0.227  & -0.408 \\ \hline
\textbf{spore, india, usa, aus, israel, uk}       & 3.153 & 2.622 & 1.957 & 1.292 & 0.627 & -0.037 & -0.702 \\ \hline
\textbf{brazil, india, sa, japan, tw, spore}      & 2.873 & 2.451 & 1.924 & 1.397 & 0.870 & 0.342  & -0.184 \\ \hline
\textbf{aus, brazil, sa, india, eu, tw, spore}  & 2.869 & 2.426 & 1.872 & 1.317 & 0.763 & 0.209  & -0.345 \\ \hline
\textbf{sa, can, china, uk, india}                  & 2.822 & 2.360 & 1.782 & 1.203 & 0.625 & 0.047  & -0.530 \\ \hline
\textbf{brazil, aus, tw, china, uk, sa, india}  & 2.789 & 2.360 & 1.824 & 1.288 & 0.752 & 0.216  & -0.319 \\ \hline
\textbf{israel, tw, japan, spore, sk, india}      & 2.735 & 2.263 & 1.673 & 1.083 & 0.493 & -0.097 & -0.687 \\ \hline
\textbf{brazil, spore, can, sa, china, india, mex} &  2.733 &  2.271 &  1.695 &  1.119 &  0.542 &  -0.033 &  -0.610 \\ \hline
\end{tabular}
\end{table}

ETFs trades are subject to commissions depending on the trading platform used. Interactive Brokers \cite{costs} for example charges \$0 commissions for trade volumes under 300,000 shares. However, commissions apply in a tiered manner for volumes higher than that. We have not factored in costs of commissions in our analysis.

\pagebreak

\section{Performance of Emerging vs Developed Economies ETFs}

As the ETFs we picked are all country specific, we applied our strategy by splitting our "master" basket into an emerging economies basket and developed economies basket.
\\
\\
We used MSCI's market classification \cite{msci} to split these ETFs as follows - \\
\textbf{Emerging Economies} - Brazil, India, South Korea, China, Mexico, South Africa and Taiwan
\\
\textbf{Developed Economies} - Japan, USA, UK, EU, Australia, Canada, Singapore and Israel.

\begin{figure}[H]
    \centering
    \caption{Sharpe Ratios of Emerging vs Developed market buckets}
    \includegraphics[scale=0.4]{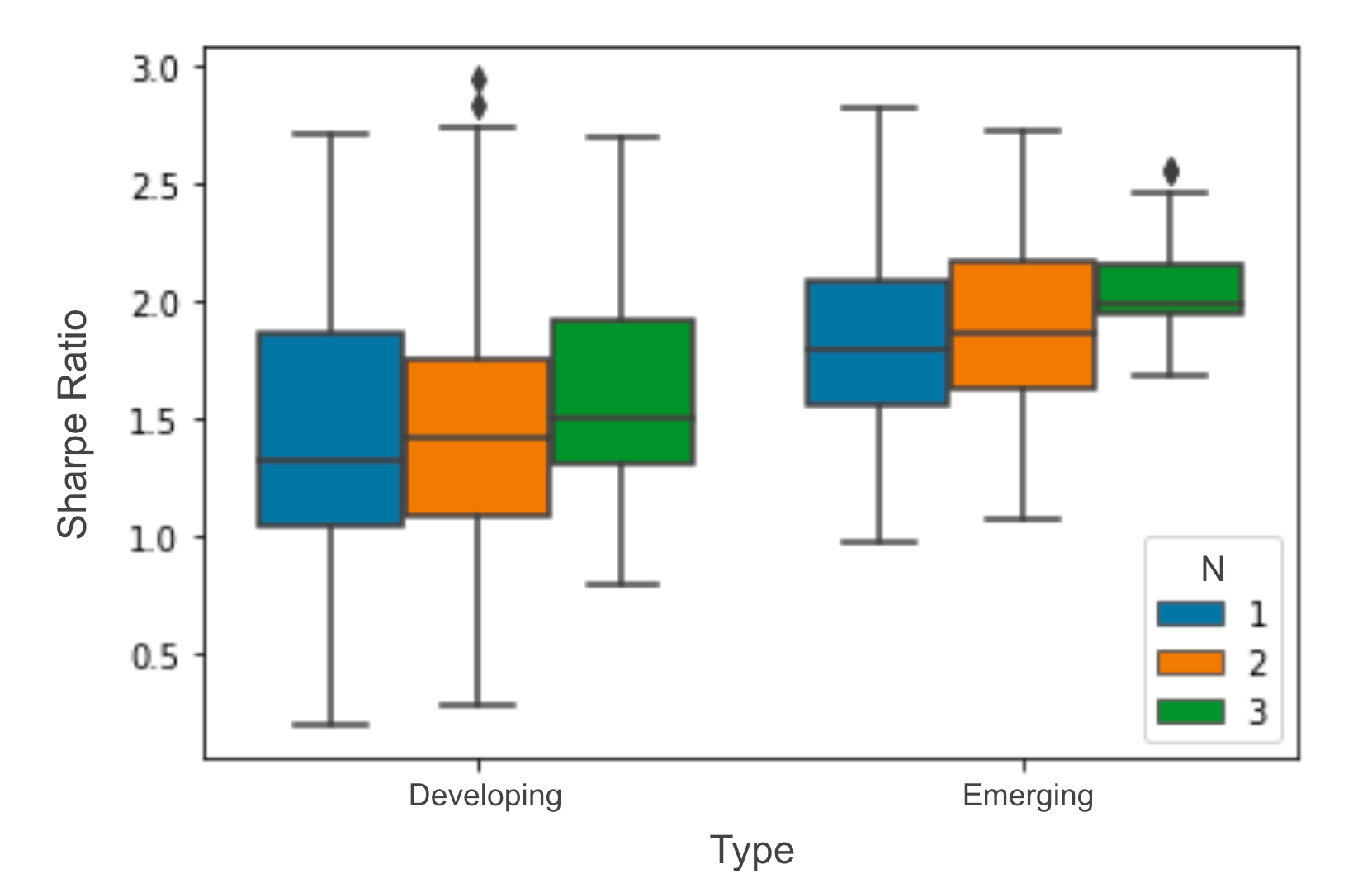}
\end{figure}

We see that while emerging market ETFs had a higher mean and median Sharpe ratio for single and multi-ETF strategies, developed market ETFs had the best performing bucket ({EU, Australia, Japan, Singapore, Israel, UK} at 2.9).

\begin{table}[htpb]
  \caption{Top 5 ETF Buckets \\
  (From Emerging + Developed Buckets Only)} 
  \label{de_basket_results}
    \begin{tabular}{|l|l|l|l|l|l|}
    \hline
    \textbf{ETF\_Basket} & \textbf{Sharpe\_Ratio} & \textbf{Type} \\ \hline
    eu, aus, japan, spore, israel, uk  & 2.942 & Developed \\ \hline
    spore, uk, israel, eu, can, aus    & 2.830 & Developed \\ \hline
    tw, sa, china, india                   & 2.819 & Emerging \\ \hline
    spore, uk, usa, israel, japan, aus & 2.733 & Developed \\ \hline
    india, tw, sa, china, sk             & 2.724 & Emerging \\ \hline
    \end{tabular}
  \end{table}
    
  \begin{table}[htpb]
  \caption{Top 5 ETF Buckets \\
  (From Emerging + Developed + Mixed Buckets)} 
  \label{mixed_basket_results}
  \begin{tabular}{|l|l|l|l|l|l|}
  \hline
  \textbf{ETF Basket} & \textbf{Sharpe Ratio} & \textbf{Type}  \\ \hline
  india, tw, can, israel, uk, spore, aus & 3.904 & Mixed \\ \hline
  india, tw, can, israel, uk, spore, aus & 3.733 & Mixed \\ \hline
  israel, tw, japan, spore, sk, india      & 3.424 & Mixed \\ \hline
  spore, india, usa, aus, israel, uk       & 3.354 & Mixed \\ \hline
  uk, india, sk, spore, china, eu          & 3.311 & Mixed \\ \hline
  \end{tabular}
  \end{table}

\begin{figure}[H]
    \centering
    \caption{Sharpe Ratios of Buckets when subjected to Constraints (Developed vs Emerging vs Mixed)}
    \includegraphics[scale=0.6]{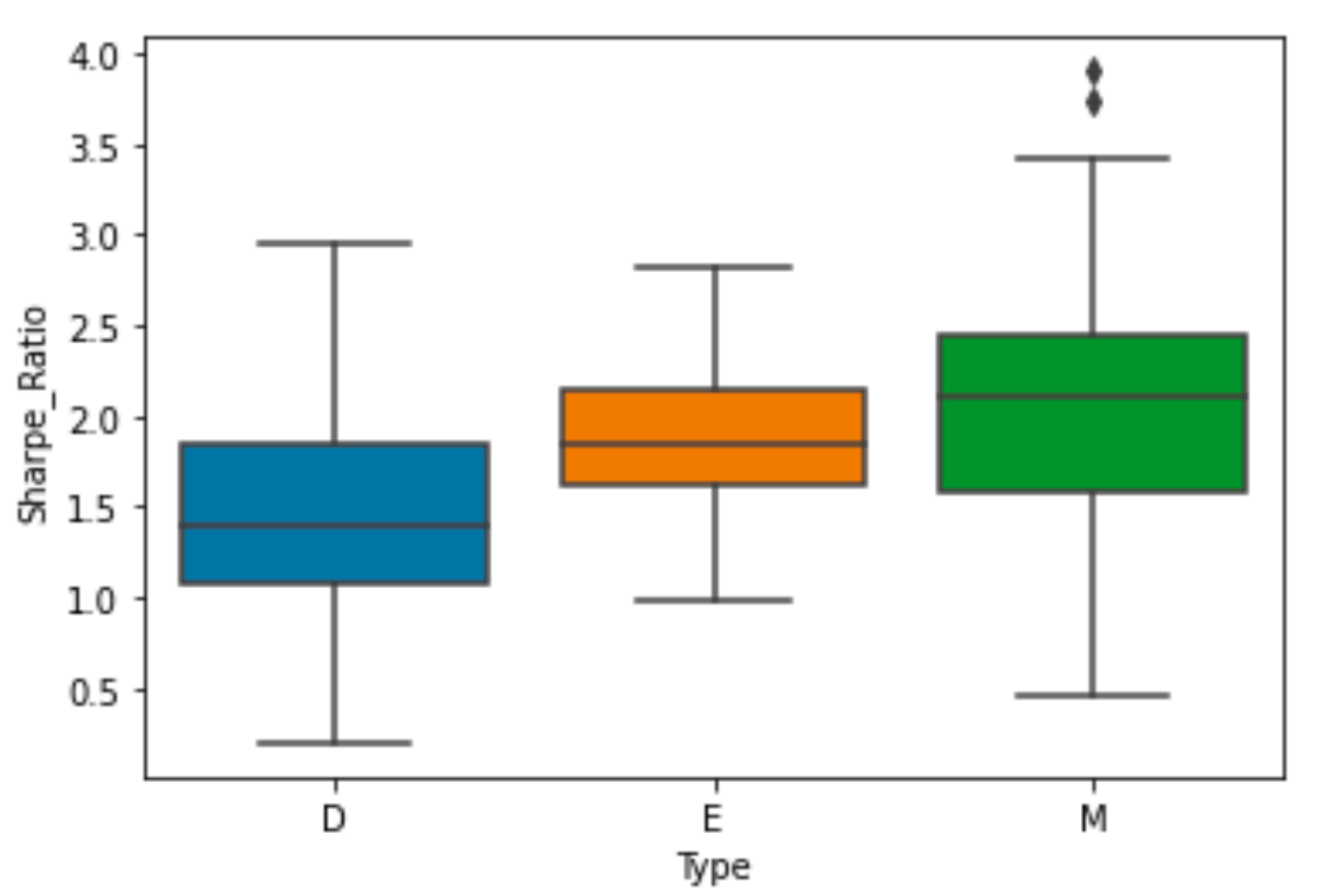}
\end{figure}
However, the overall performance is worse compared to mixed buckets. As seen in Table \ref{de_basket_results}, the ETFs when chosen from \textbf{only} Emerging or \textbf{only} Developed bucket do worse than when there is no such restriction on picking ETFs (as seen in Table \ref{mixed_basket_results}).

\pagebreak

\section{Conclusion}

We attempt to build on the Pagonidis' findings \cite{effect}. Initially, we corroborate their findings on likelihood of mean reversion based on IBS thresholds in \textbf{Table \ref{positive_prob}} on recent data.

The analysis reiterates that IBS is a strong predictor of close-to-close returns on equity ETFs. We present a new strategy that aims to maximise the IBS effect by combining ETFs in a basket and going long and short on ETFs with maximum and minimum IBS values.

The analysis includes modifications of the above strategy where aspects such as - holding period, number of held ETFs, IBS calculation periods - among others are varied.

The presented strategy also outperforms the threshold-based strategy on multiple metrics largely due to having a 100\% time-in. Our approach to use a bucket of ETFs also reduces the risk of being over-reliant on any one ETF as the signal is calculated daily and by picking more than one ETF per day, the returns are averaged between them.

We must note that while our proposed strategy does well by going long and short on various ETFs, any trading costs including Commissions, Borrow Rate and Slippage could reduce the performance of the strategy - especially on the Short Side \cite{lend-etfs}. While we have tried to include some of these costs while calculating and presenting our results, real world costs might vary.

\bibliographystyle{ieeetr}
\bibliography{sn-bibliography}

\end{document}